\documentclass[reprint,twocolumn,aps,prb,superscriptaddress,a4paper,nofootinbib,showpacs,showkeys,lengthcheck]{revtex4-1}

\setcitestyle{numbers,square}

\usepackage{graphicx}
\usepackage{type1cm}
\usepackage[normalem]{ulem}

\usepackage{bm,dcolumn,amssymb,amsmath,braket,siunitx,color}

\usepackage{afterpage}

\newcommand{\mm}[1]{\mbox{$#1$}}
\newcommand{\dstd}{\mathrm{d}}

\newcommand{\ms}{\mbox{$\mu_{\mathrm{spin}}$}}

\newcommand{\qq}{\mbox{$\bm{q}$}}

\newcommand{\epsq}{\mbox{$\epsilon(\bm{q})$}}

\newcommand{\ea}{{\it et al.}}

\definecolor{mojezel}{rgb}{0,0.5,0.2}

\begin{document}

\title{Assessing different approaches to ab~initio calculations of
  spin wave stiffness} 



\author{O. \surname{\v{S}ipr}} 
\email{sipr@fzu.cz}
\homepage{http://crysa.fzu.cz/ondra} \affiliation{FZU -- Institute of
  Physics ASCR, Cukrovarnick\'{a}~10, CZ-162~53~Prague, Czech Republic
} \affiliation{New Technologies Research Centre, University of West
  Bohemia, Pilsen, Czech Republic}

\author{S. \surname{Mankovsky}} \affiliation{Universit\"{a}t
  M\"{u}nchen, Department Chemie, Butenandtstr.~5-13,
  D-81377~M\"{u}nchen, Germany}

\author{H. \surname{Ebert}} \affiliation{Universit\"{a}t M\"{u}nchen,
  Department Chemie, Butenandtstr.~5-13, D-81377~M\"{u}nchen, Germany}

\date{\today}

\begin{abstract}
Ab~initio calculations of the spin wave stiffness constant $D$ for
elemental Fe and Ni performed by different groups in the past have led
to values with a considerable spread of 50--100~\%.  We present
results for the stiffness constant $D$ of Fe, Ni, and permalloy
Fe$_{0.19}$Ni$_{0.81}$ obtained by three different approaches: (i) by
finding the quadratic term coefficient of the power expansion of the
spin wave energy dispersion, (ii) by a damped real-space summation of
weighted exchange coupling constants, and (iii) by integrating the
appropriate expression in reciprocal space. All approaches are
implemented by means of the same Korringa-Kohn-Rostoker (KKR) Green
function formalism.  We demonstrate that if properly converged, all
procedures yield comparable values, with uncertainties of 5--10~\%
remaining.  By a careful analysis of the influence of various
technical parameters we estimate the margin of errors for the
stiffness constants evaluated by different approaches and suggest
procedures to minimize the risk of getting incorrect results.
\end{abstract}



\maketitle


\flushbottom


\section{Introduction}   \label{sec-intro}

Investigations of magnetic properties of materials at the
phenomenological level based on a model spin Hamiltonian may be
formulated either in a continuous field (micromagnetic) representation
or in an atomistic representation.  In the first case, the energy
functional (neglecting relativistic effects) is given by
\cite{VBL+08}
\begin{equation}
E[\bm{m}] \: = \: \int_{V} \!
\dstd^{3} \bm{r} \, A_{\text{ex}} 
\sum_{c=x,y,z}
\left(\frac{\partial \bm{m}}{\partial c}\right)^{2}
\; ,
\label{eq-ex}
\end{equation}
where \mm{\bm{m}(\bm{r})} is the magnetization field and
$A_{\text{ex}}$ is the exchange stiffness constant. In the atomistic
representation, the energy is given by the Heisenberg Hamiltonian,
\begin{equation}
 H \: = \: - \, \sum_{ij} J_{ij} \,
 \bm{\hat{e}}_{i} \cdot  \bm{\hat{e}}_{j}
 \; ,
 \label{eq-heisen}
\end{equation}
where~$\bm{\hat{e}}_{i}$ and ~$\bm{\hat{e}}_{j}$ are unit vectors
specifying the orientation of the magnetic moments for the atoms $i$
and $j$, with the exchange coupling characterized by the exchange
parameter $J_{ij}$. Both approaches give access to the energy of spin
wave excitations~$\epsilon(\bm{q})$, which can be written in the long
wave limit as \cite{Kub+00}
\begin{equation}
\epsilon(\bm{q}) \; = \;
  D \, |\bm{q}|^{2} \: + \: \ldots
\; ,
\label{eq-D}
\end{equation}
where~$\bm{q}$ is is the corresponding wave vector and $D$ is the spin
wave stiffness constant.  The quantity is directly connected to the
exchange stiffness $A_{\text{ex}}$ via the relation \cite{VBL+08}
\begin{equation}
  A_{\text{ex}} \; = \; \frac{ D \, M_{s}}{ 2 g \mu_{B}}
  \; ,
\label{eq-A-from-D}
\end{equation}
where $M_{s}$ is the saturation magnetization, $g$ is the Land\'{e}
factor ($g \approx 2$ for metals) and~$\mu_{B}$ is the Bohr magneton.
This relation provides a link to the experiment: the model parameter
$A_{\text{ex}}$ entering Eq.~(\ref{eq-ex}) can be obtained from the
spin wave stiffness constant $D$, which can be determined
experimentally.  On the other hand, comparing the spin wave stiffness
calculated from first principles with experimental data allows to
assess the reliability of models and approximations used in the
calculations.

As the spin wave stiffness constant $D$ characterizes the energy of
spin-wave excitations in the long-wave limit, it can be obtained on
the basis of spin spiral calculations within the adiabatic
approximation. In this case the energy \epsq\ calculated from first
principles for a spin spiral characterized by a wave vector
\qq\ should be approximated around $\bm{q}$=0 by a suitable polynomial
and the stiffness constant $D$ is just the expansion coefficient of
the quadratic term.

A similar scheme can be applied also to evaluate $D$ by using the
spin-wave energy dispersion represented in terms of real-space
interatomic exchange coupling parameters, obtained on the basis of the
magnetic force theorem. In this case, by expressing the spin spiral
energy in a power series of \qq, one arrives at the expression
\cite{LKG84,LKAG87,PKT+01}
\begin{equation}
  D \: =\: \sum_{j} 
  \frac{2 \mu_{B}}{3 \mu_{j}} \,
  J_{0j} \, R_{0j}^{2}
\; ,
  \label{eq-simple}
\end{equation}
where~$\mu_{j}$ is the magnetic moment of atom $j$ and $R_{0j}$ is the
corresponding inter-atomic distance.

\begin{table}
\caption{Previous theoretical results for the spin wave stiffness $D$
  (in meV~\AA$^{2}$) of elemental Fe and Ni and of Fe$_{1-x}$Ni$_{x}$
  alloy
  (its composition was Fe$_{0.25}$Ni$_{0.75}$ \cite{YJK+08} and
    Fe$_{0.19}$Ni$_{0.81}$ \cite{PCD+17}). 
  Each study is identified by a reference and the publication year.
  The method how $D$ was evaluated is indicated in the last column:
  $\epsilon(\bm{q})$ stands for fitting the spin wave dispersion
  Eq.~(\ref{eq-D}) whereas $J_{ij}$ denotes a weighted sum of the
  coupling constants Eq.~(\ref{eq-simple}).}
\label{tab-d-the}
\begin{ruledtabular} 
\begin{tabular}{rcccl}
  \multicolumn{1}{c}{work}  &  Fe & Ni &  Fe$_{1-x}$Ni$_{x}$  &  method \\
\hline
\cite{MFL+96}\hspace{1em}(1996)       &
214     & 527     &  & $J_{ij}$ \\
\cite{RJ+97}\hspace{1em}(1997)    &
247     & 739     &  & $J_{ij}$ \\
\cite{BNW+99}\hspace{1em}(1999)         &
135     & 180     &  &  $\epsilon(\bm{q})$  \\
\cite{SA+99}\hspace{1em}(1999) &
280     & 740     &   & $\epsilon(\bm{q})$ \\
\cite{Kub+00}\hspace{1em}(2000)    &
355  & 790  &  & $\epsilon(\bm{q})$ \\
\cite{PKT+01}\hspace{1em}(2001)         &
250  & 756  &  & $J_{ij}$ \\
\cite{MEF+03}\hspace{1em}(2003)         &
200  &  &   & $\epsilon(\bm{q})$ \\
\cite{YJK+08}\hspace{1em}(2008) &
  &  &  515  & $J_{ij}$ \\
\cite{SKMR05}\hspace{1em}(2005)    &
322 & 541 &   & $\epsilon(\bm{q})$ \\
\cite{PCD+17}\hspace{1em}(2017)           &
320 & 707 & 620 & $J_{ij}$  \\
\end{tabular}
\end{ruledtabular}
\end{table}

Both approaches, i.e., the one based on fitting calculated spin spiral
energies by a polynomial and the one based on evaluating the
real-space sum Eq.~(\ref{eq-simple}), were employed for {\em ab
  initio} calculations of the spin wave stiffness constant $D$ in the
past.  However, despite the conceptual simplicity of both procedures,
values of $D$ obtained by different groups for the same systems
exhibit a considerable spread.  To illustrate this, we present in
Tab.~\ref{tab-d-the} several values for the stiffness constant $D$ for
elemental Fe and Ni and permalloy Fe$_{0.19}$Ni$_{0.81}$ (Py) obtained
by previous theoretical studies.  Another comparison can be found,
e.g., in Table~10 of Vaz \ea\ \cite{VBL+08}. One can see from
Tab.~\ref{tab-d-the} that the deviations may easily reach 50\%.  Such
big discrepancies are extraordinarily high when compared, for example,
with the situation for the exchange coupling parameters $J_{ij}$ ---
even when considering that the studies employ different methods of
electronic structure calculations relying on different approximations.
The discrepancies appear also between studies which use the same
method to evaluate the stiffness constant $D$: it is not that results
based on one method would cluster around one value and results based
on the other method around another value.  As none of the previous
studies presented results obtained by both methods --- one always
focused solely either on Eq.~(\ref{eq-D}) or on Eq.~(\ref{eq-simple})
--- it is difficult to assess properly the accuracy and reliability of
the procedures involved.

Yet another way to determine the spin-wave stiffness as the second
derivative of the spin-wave energy with respect to the wave vector
$\bm{q}$ relies on evaluating the corresponding derivatives of the
exchange parameters $J(\bm{q})$. This in turn leads to an expression
for the stiffness constant $D$ formulated in reciprocal space, as
presented by Liechtenstein \ea\ \cite{LKG84,LKAG87}.  So far no
results on spin-wave stiffness based on this third approach have been
reported in the literature. Only recently, a relativistic extension of
the reciprocal-space expression for $D$ was presented by Mankovsky
\ea\ \cite{MPE+19} and applied for studying Fe-Ni alloys.

The stiffness constant $D$ is an important characteristic quantity as
it determines --- together with the magnetic anisotropy --- the domain
structure of magnetic materials.  In addition, it determines the
magnetization dynamics.  Therefore, there is an urgent need for a
detailed comparison of the various computational approaches to
evaluate $D$ so that reliable values can be obtained.

The aim of this work is therefore to calculate the spin wave stiffness
constant $D$ of Fe, Ni, and Py by analyzing the long-wave limit of the
spin wave dispersion relation Eq.~(\ref{eq-D}), by evaluating the
weighted sum of coupling constants Eq.~(\ref{eq-simple}), and by a
direct evaluation of $D$ in reciprocal space \cite{MPE+19}.  All three
approaches are implemented by use of the same electronic structure
method, namely, the Korringa-Kohn-Rostoker (KKR) Green function
formalism, meaning that the results are directly comparable.  A
careful analysis of the influence of various technical parameters
makes it possible to estimate the margin of errors for the stiffness
constants evaluated by the different approaches and to decide whether
there is a significant difference between them or not.  By
illustrating how various factors affect the outcome, we offer a
guidance how the procedures ought to be performed to minimize the
risks of wrong results.



\section{Computational scheme}   \label{sec-comput}

Evaluation of the spin wave stiffness constant $D$ is done on the
basis of a calculation of the underlying electronic structure of the
system.  For this we employed the {\em ab~initio} spin-polarized
multiple-scattering or Korringa-Kohn-Rostoker (KKR) Green function
formalism \cite{EKM11} as implemented in the {\sc sprkkr} code
\cite{sprkkr-code}.  The calculations were performed in a
scalar-relativistic mode, relying on the generalized gradient
approximation (GGA) to the spin density functional theory, using the
Perdew, Burke and Ernzerhof (PBE) functional.  For the multipole
expansion of the Green function, an angular momentum cutoff
\mm{\ell_{\mathrm{max}}}=3 was used.  The potentials were subject to
the atomic sphere approximation (ASA).  When dealing with Py, the
substitutional disorder was accounted for within the coherent
potential approximation (CPA).  The energy integrals were evaluated by
contour integration on a semicircular path within the complex energy
plane, using a Gaussian mesh of 32~points.  The $\bm{k}$-space
integration was carried out via sampling on a regular mesh, making use
of the symmetry.  The number of $\bm{k}$-points in the mesh is an
important technical parameter and will be considered in
Sec.~\ref{sec-res} in more details; here we just note that unless
specified otherwise, we used $89^{3}$ points in the full Brillouin
zone (BZ) for bcc Fe, $112^{3}$ points for fcc Ni and $75^{3}$ points
for fcc Py.  The equilibrium lattice constant $a_{0}$ was determined
by minimizing the total energy for each system.  This gives us
$a_{0}=2.830$\AA\ for Fe, 3.506\AA\ for Ni, and 3.523\AA\ for Py.

To evaluate the stiffness constant $D$ by means of finding the
expansion coefficient as in Eq.~(\ref{eq-D}), the spin wave energy
dispersion relation $\epsilon(\bm{q})$ has to be obtained.  We
achieved this by evaluating the change of the total energy per unit
cell $E(\bm{q},\theta)-E(0,\theta)$ due to a spin spiral characterized
by magnetic moment
\begin{equation}
\mu_{\text{spin}} \, [ \,
  \cos(\bm{q}\bm{R}) \, \sin\theta \, , \,
  \sin(\bm{q}\bm{R}) \, \sin\theta \, , \,
  \cos\theta \, ]
\; ,
\label{eq-msp}
\end{equation}
where~$\bm{R}$ is the Bravais lattice vector,  $\bm{q}$ is the spin
spiral wave vector,  $\theta$ is the spiral cone angle
 and \ms\ is the magnitude of the magnetic moment per
site.  The magnon energy  $\epsilon(\bm{q})$ is given by 
\cite{Kub+00}
\begin{equation}
  \epsilon(\bm{q}) \: = \:
  \lim_{\theta \rightarrow 0} \frac{4 \mu_{B}}{\mu_{\text{spin}}}
  \frac{E(\bm{q},\theta)-E(0,\theta)}{\sin^{2}\theta}
  \; .
\label{eq-ene}
\end{equation}
The change of the energy $E(\bm{q},\theta)-E(0,\theta)$ due to the
spin spiral Eq.~(\ref{eq-msp}) can be obtained either by employing
self-consistent calculations for each of the wave vectors~$\bm{q}$ or
by relying on the force theorem, meaning that the same potential
(obtained for the ferromagnetic state) for all wave vectors~$\bm{q}$
is used.  The electronic structure for spiral magnetic order was
calculated as described by Mankovsky \ea\ \cite{MFE11}.
 Our calculations are scalar-relativistic, therefore, the results
  do not depend on the angle between the axis of the spin rotation
  cone and the spin wave propagation direction. 
The details
how the behavior of $\epsilon(\bm{q})$ for~$\bm{q}\rightarrow0$ was
analyzed are described in detail in Sec.~\ref{sec-spir}.

When resorting to the second option, namely, an evaluation of the
stiffness constant $D$ in real space by relying on
Eq.~(\ref{eq-simple}), one has to deal with the fact that the sum over
the atoms $\sum_{j}$ in Eq.~(\ref{eq-simple})   does not converge for
metals (due to the long-range character of the exchange coupling)
\cite{PKT+01}.  
Hence, an additional damping factor has been introduced
which enables evaluation of Eq.~(\ref{eq-simple}) by extrapolating the
partial results to zero damping \cite{PKT+01}.  In particular we
evaluated the stiffness constant $D$ as \cite{PKT+01,TCF+09,DGC+15}
\begin{align}
  D &= \lim_{\eta \rightarrow 0} D(\eta)
  \; ,  \label{eq-dlim} \\
  D(\eta) &= \sum_{\alpha} c_{\alpha} \, D_{\alpha}(\eta)
  \; ,  \label{eq-dalpha} \\
  D_{\alpha}(\eta) &=  \sum_{j} \sum_{\beta} \,
  c_{\beta} \, 
  \frac{2 \mu_{B}}{3\sqrt{|\mu_{\alpha}| |\mu_{\beta}|}} \,
  J_{0j}^{(\alpha\beta)} \, R_{0j}^{2} \,
  \mathrm{e}^{-\eta \frac{R_{0j}}{R_{01}}}
  \; ,
   \label{eq-dsum}
\end{align}
where $j$ labels the lattice sites, $c_{\alpha}$ and~$\mu_{\alpha}$
are the concentration and the magnetic moment of atoms of type
$\alpha$, $J_{0j}^{(\alpha\beta)}$ is the pairwise exchange coupling
constant if an atom of type~$\alpha$ is located at the lattice origin
and an atom of type~$\beta$ is located at the lattice site $j$,
$R_{0j}$ is the distance of the site $j$ from the lattice origin,
$\eta$ is the damping parameter and $R_{01}$ is the nearest-neighbor
interatomic distance \cite{PKT+01,TCF+09,DGC+15,SME+19}.  Note that we
have one atomic type for Fe and Ni whereas two atomic types for Py.
The exchange coupling constants $J_{0j}^{(\alpha\beta)}$ were
evaluated from the electronic structure using the prescription of
Liechtenstein \ea\ \cite{LKAG87}.  Taking the limit $\lim_{\eta
  \rightarrow 0} D(\eta)$ in Eq.~(\ref{eq-dlim}) is a delicate issue
and we devote to it most of Sec.~\ref{sec-jij}.

Performing the sum over atomic sites in Eqs.~(\ref{eq-simple}) or
(\ref{eq-dalpha})--(\ref{eq-dsum}) can be by-passed by evaluating the
stiffness constant $D$ via a reciprocal-space integration
\cite{LKG84,LKAG87}.  A recently reported relativistic generalization
of this approach \cite{MPE+19} leads to the expression:
\begin{widetext}
\begin{equation}
D_{\alpha\beta} \: = \:
\frac{1}{\pi\mu_{\text{spin}}} \mbox{Im} \mbox{Tr} \,
\int^{E_F} \! \dstd E \:
\frac{1}{\Omega_{\text{BZ}}} \int_{\text{BZ}} \! \dstd^{3}k \:
\left[
  \underline{T}_{x}
  \frac{\partial \underline{\tau}(\bm{k},E)}{\partial k_{\alpha}}
  \underline{T}_{x}
  \frac{\partial \underline{\tau}(\bm{k},E)}{\partial k_{\beta}}
  \, + \,
  \underline{T}_{y}
  \frac{\partial \underline{\tau}(\bm{k},E)}{\partial k_{\alpha}}
  \underline{T}_{y}
  \frac{\partial \underline{\tau}(\bm{k},E)}{\partial k_{\beta}}
  \right]
\; .
\label{eq-sergey}
\end{equation}
\end{widetext}
The matrix~$\underline{\tau}$ is the Fourier transform of the
scattering path operator and the matrices~$\underline{T}_{x}$,
$\underline{T}_{y}$ represent the change of the potential upon
rotating the spin,
\begin{align}
  T_{x,\Lambda_{1}\Lambda_{2}} \, =& \,
  \int \!  \dstd^{3}r \, Z^{\times}_{\Lambda_{1}}(\bm{r},E) \,
  \beta \sigma_{x} \,
  Z_{\Lambda_{2}}(\bm{r},E)
  \; , \\
  T_{y,\Lambda_{1}\Lambda_{2}} \, =& \,
  \int \!  \dstd^{3}r \, Z^{\times}_{\Lambda_{1}}(\bm{r},E) \,
  \beta \sigma_{y} \,
  Z_{\Lambda_{2}}(\bm{r},E)
  \; ,
\end{align}
where $Z_{\Lambda}(\bm{r},E)$ stands for the regular solution of a
single-site Dirac equation, and subscripts~$\alpha$, $\beta$ denote
cartesian components.  We deal with cubic lattices, so the stiffness
constant $D$ is isotropic, $D_{xx}=D_{yy}=D_{zz}=D$.  For more
details, see the original paper \cite{MPE+19}.  The advantage of using
Eq.~(\ref{eq-sergey}) is that there is no need for polynomial fitting
as when employing Eqs.~(\ref{eq-D}) and (\ref{eq-ene}) or for
extrapolation as when employing Eq.~(\ref{eq-dlim}).  On the other
hand, when proceeding along Eq.~(\ref{eq-sergey}), one has to evaluate
the derivative $(\partial \tau/\partial k)$ which is a numerically
demanding task.

\raggedbottom

In this work we evaluated the integrand in Eq.~(\ref{eq-sergey}) using
the same scalar-relativistic potential as when obtaining the stiffness
constant $D$ via Eqs.~(\ref{eq-D}) and (\ref{eq-ene}) or via
Eqs.~(\ref{eq-dlim})--(\ref{eq-dsum}).  Moreover, we suppress the
spin-orbit coupling (SOC) by employing an approximate two-component
scheme \cite{EFVG96}, similarly as when investigating the influence of
SOC on electronic-structure-related properties in the past
\cite{SBE+14,SMP+16}.  The results we obtain by means of
Eq.~(\ref{eq-sergey}) are thus directly comparable to
scalar-relativistic results obtained by means of the other two
approaches.  The~$\bm{k}$-mesh used for this type of calculations
contained $135^3$ points for Fe and Py and $144^3$ points for Ni.

 Let us note finally that even though we employ a particular
  electronic structure calculation method (KKR Green function
  formalism), the issues we deal with are not specific to it and will
  have to be cared upon no matter which calculational method is used.

\makeatletter
\afterpage{\global\let\@textbottom\relax \global\let\@texttop\relax}



\section{Results}   \label{sec-res}



\subsection{Fitting spin wave energy dispersion}  

\label{sec-spir}

 First we consider various aspects when obtaining the spin-wave
  stiffness constant $D$ as the coefficient of the quadratic term
  of the power expansion for the spin-wave energy
  dispersion relation Eq.~(\ref{eq-D}).  An obvious technical
  parameter against which the convergence should be checked is the
  density of the mesh used to evaluate the integrals in
  $\bm{k}$-space.  We verified that for the grids used in this
  section, namely, $165^{3}$ points in the full Brillouin zone for Fe,
  to $149^{3}$ points for Ni, and to $133^{3}$ points for Py, the
  values of $D$ are converged within $\pm$0.2~meV~\AA$^{2}$.  This
  means that for the purpose of the convergency tests outlined in this
  section, the values of $D$ can be considered as practically
  accurate.

\begin{figure}
\includegraphics[viewport=0.2cm 0.5cm 8.7cm 11.0cm]{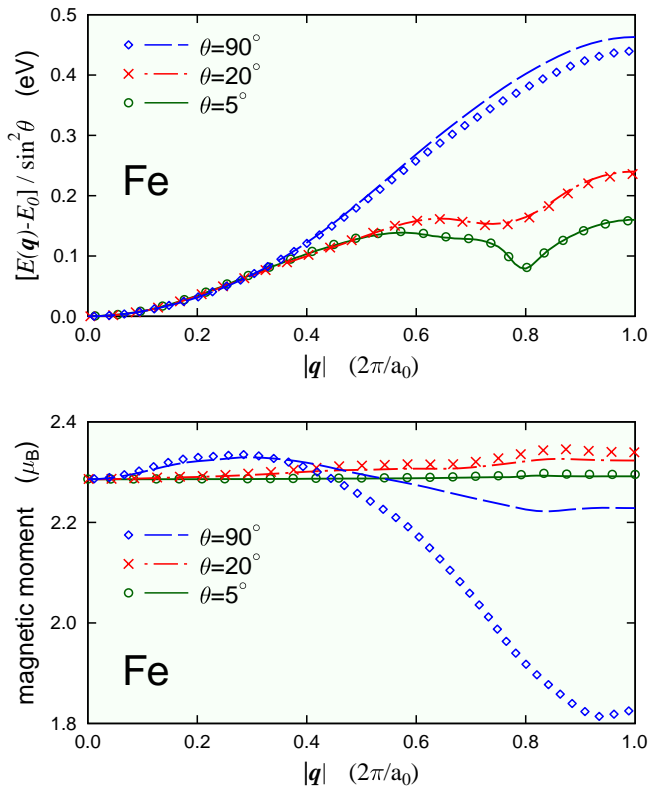}%
\caption{(Color online) Magnetic moment per atom (lower panel) and
  energy dispersion \mm{[E(\bm{q})-E_{0}]/\sin^{2}\theta}\ (upper
  panel) for spin spiral waves propagated along the [001]
  direction in Fe, obtained by means of self-consistent calculations
  (markers) and by means of magnetic force theorem (lines).  The
  spiral cone angle~$\theta$ is specified in the legend. }
\label{fig-eneq-fe}
\end{figure}

\begin{figure}
\includegraphics[viewport=0.2cm 0.5cm 8.7cm 11.0cm]{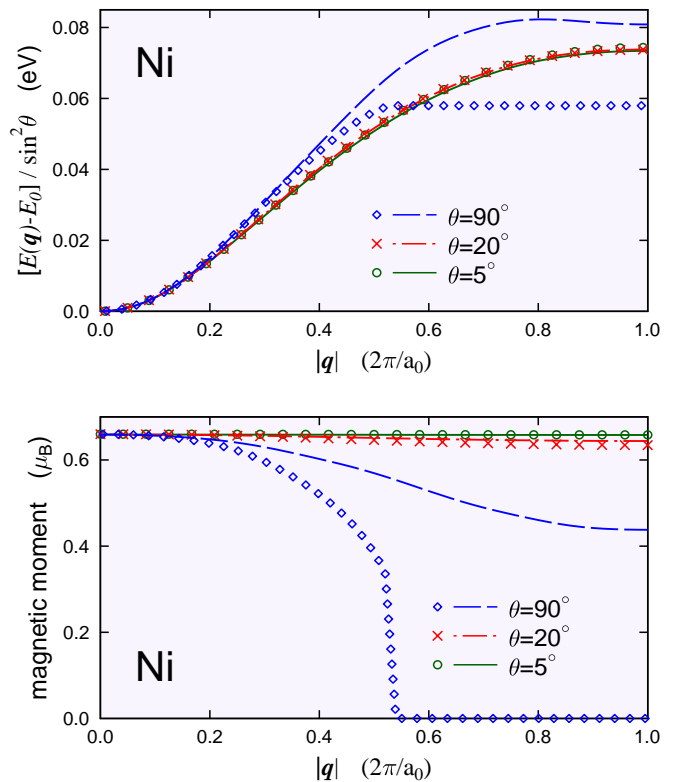}%
\caption{(Color online) As Fig.~\protect\ref{fig-eneq-fe} but for Ni.
Note that the energy dispersion curves for $\theta=5^{\circ}$ and
$20^{\circ}$ almost coincide.}
\label{fig-eneq-ni}
\end{figure}

 The magnon energy  $\epsilon(\bm{q})$ is represented 
in terms of the spin-spiral energy Eq.~(\ref{eq-ene}) in the limit
\mm{\theta \rightarrow 0}.
For this one should calculate the electronic
structure for spin spirals with the cone angle~$\theta$ as small as
possible.  However, if the angle~$\theta$ approaches zero, so does the
energy difference \mm{E(\bm{q},\theta)-E(0,\theta)}, and evaluating
the ratio \mm{[E(\bm{q},\theta)-E(0,\theta)]/\sin^{2}\theta}\ becomes
numerically unstable.  Therefore, we start by looking closely on the
sensitivity of \mm{[E(\bm{q},\theta)-E(0,\theta)]/\sin^{2}\theta}\ to
the value of~$\theta$.

To get an overview, we plot the ratio
\begin{equation}
  \frac{E(\bm{q},\theta) - E(0,\theta)}{\sin^{2}\theta}
    \label{eq-elim}
\end{equation} 
as a function of the wave vector~$\bm{q}$ for several values of the
cone angle~$\theta$.  This is done in the upper panels of
Figs.~\ref{fig-eneq-fe} and \ref{fig-eneq-ni} for Fe and Ni,
respectively.  The wave vector~$\bm{q}$ is oriented along the [001]
direction.  The corresponding spin spiral energies were calculated by
means of self-consistent calculations (SCF), i.e., with the potential
recalculated for each~$\bm{q}$ vector (shown via markers), as well as
using the magnetic force theorem (MFT), with the potential taken
always the same as for $\bm{q}$=0 (shown via lines).  Additionally, we
present data on magnetic moments in the lower panels of
Figs.~\ref{fig-eneq-fe}--\ref{fig-eneq-ni}, to provide a more complete
picture.  Interestingly, the magnetic moment is more sensitive to
whether the calculation is done self-consistently or not than the
energy is --- especially if $|\bm{q}|$ gets large. The sudden decrease
of \ms\ of Ni at about \mm{0.5(2\pi/a_{0})} clearly seen for
\mm{\theta=90^{\circ}}\ (Fig.~\ref{fig-eneq-ni}) corresponds to the
well-known collapse of the magnetic moment of Ni in case of
anti-ferromagnetic order \cite{THO+82,RJ+97}.

\begin{table}
\caption{Spin wave stiffness constant $D$ of Fe evaluated by fitting
  the energy dispersion Eq.~(\ref{eq-theta}) for~\mm{\bm{q} \rightarrow
    0}\ by a bi-quadratic polynomial Eq.~(\ref{eq-biq}), for spiral cones angles
  \mm{\theta=20^{\circ}}, $45^{\circ}$, and $90^{\circ}$.  The
  energies were obtained either from self-consistent calculations
  (SCF) or by employing the magnetic force theorem (MFT). }
\label{tab-spir-D-fe}
\begin{ruledtabular}
\begin{tabular}{ccc}
   &   \multicolumn{1}{c}{SCF}
  &    \multicolumn{1}{c}{MFT} \\
  \multicolumn{1}{c}{\mbox{$\theta$~(${}^{\circ}$)}} &
  \multicolumn{1}{c}{\mbox{$D$ (meV~\AA$^{2}$)}} &
  \multicolumn{1}{c}{\mbox{$D$ (meV~\AA$^{2}$)}} \\
\hline
20  &  302.4  &   292.8 \\
45  &  301.1  &   292.3 \\
90  &  301.7  &   294.0 \\
\end{tabular}
\end{ruledtabular}
\end{table}

\begin{table}
\caption{Same as Tab.~\protect\ref{tab-spir-D-fe} but for Ni. }
\label{tab-spir-D-ni}
\begin{ruledtabular}
\begin{tabular}{ccc}
   &   \multicolumn{1}{c}{SCF}
  &    \multicolumn{1}{c}{MFT} \\
  \multicolumn{1}{c}{\mbox{$\theta$~(${}^{\circ}$)}} &
  \multicolumn{1}{c}{\mbox{$D$ (meV~\AA$^{2}$)}} &
  \multicolumn{1}{c}{\mbox{$D$ (meV~\AA$^{2}$)}} \\
\hline
20  &  752.4  &  747.9  \\
45  &  753.4  &  746.9  \\
90  &  755.7  &  745.9  \\
\end{tabular}
\end{ruledtabular}
\end{table}

\begin{table}
\caption{Same as Tab.~\protect\ref{tab-spir-D-fe} but for Py. }
\label{tab-spir-D-py}
\begin{ruledtabular}
\begin{tabular}{ccc}
   &   \multicolumn{1}{c}{SCF}
  &    \multicolumn{1}{c}{MFT} \\
  \multicolumn{1}{c}{\mbox{$\theta$~(${}^{\circ}$)}} &
  \multicolumn{1}{c}{\mbox{$D$ (meV~\AA$^{2}$)}} &
  \multicolumn{1}{c}{\mbox{$D$ (meV~\AA$^{2}$)}} \\
\hline
20  &  519.7  &  520.6  \\
45  &  521.6  &  522.9  \\
90  &  521.2  &  522.2  \\
\end{tabular}
\end{ruledtabular}
\end{table}

Based on the curves in Figs.~\ref{fig-eneq-fe}--\ref{fig-eneq-ni}, it
appears that for $|\bm{q}|$ less than about \mm{0.2(2\pi/a_{0})}, the
ratio Eq.~(\ref{eq-elim}) depends only little on~$\theta$ and that the
differences between SCF and MFT calculations are small.  To get more
quantitative information on the dependence of the expression in
Eq.~(\ref{eq-elim}) on the cone angle~$\theta$, we summarize in
Tabs.~\ref{tab-spir-D-fe}--\ref{tab-spir-D-py} the values of $D$
obtained by fitting a bi-quadratic function
\begin{equation}
  f_{4}(q,\theta) \: = \:
  a_{2}(\theta) q^{2} \: + \: a_{4}(\theta) q^{4} \; ,
  \label{eq-biq}
\end{equation}
to the energy
\begin{equation}
  \epsilon(\bm{q},\theta) \: = \:
  \frac{4 \mu_{B}}{\mu_{\text{spin}}}
  \frac{E(\bm{q},\theta)-E(0,\theta)}{\sin^{2}\theta}
\label{eq-theta}
\end{equation}
in the interval $|\bm{q}|\in$[0,$0.1(2\pi/a_{0})$], for different
values of the cone angle~$\theta$.  The stiffness constant $D$ is
obtained as the expansion coefficient of the quadratic term,
\[
D \, = \, a_{2}(\theta)  \; .
\]
The necessary $\bm{k}$-space integrals were evaluated using a regular
mesh corresponding to $165^{3}$ points in the full BZ for Fe, to
$149^{3}$ points for Ni, and to $133^{3}$ points for Py.  The results
based on SCF and MFT calculations are shown separately.
The~$\bm{q}$-vector was varied along the [001] direction, the fit was
obtained for values of $|\bm{q}|$ from zero to \mm{0.1(2\pi/a_{0})}.
Data for $\theta<20^{\circ}$ are not included in
Tabs.~\ref{tab-spir-D-fe}--\ref{tab-spir-D-py} because for small
values of~$\theta$ the stability of the fit gets worse due to very
small differences $E(\bm{q},\theta)-E(0,\theta)$ and the results are
not reliable.

One can infer from Tabs.~\ref{tab-spir-D-fe}--\ref{tab-spir-D-py} that
going to very low values of~$\theta$ is not needed for spin wave
stiffness calculations.  The energy $\epsilon(\bm{q},\theta)$ is
practically independent on the cone angle~$\theta$.  Consequently, all
the spin spiral calculations are done for \mm{\theta=20^{\circ}} in
the rest of the paper, because this value appears to be a good
representation of \mm{\theta \rightarrow 0}\ for the purpose of
evaluating Eq.~(\ref{eq-ene}) and still is large enough to lead to
numerically stable results.

\begin{figure}
\includegraphics[viewport=0.2cm 0.5cm 8.7cm 5.9cm]{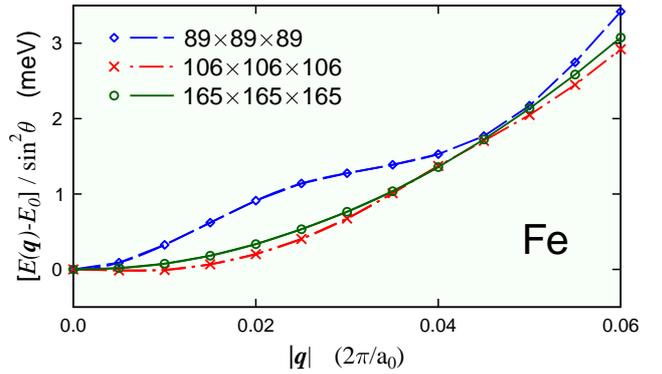}%
\caption{(Color online) The energy dispersion
  \mm{[E(\bm{q})-E_{0}]/\sin^{2}\theta}\ for spin spiral waves in Fe
  obtained when the $\bm{k}$-space integrals were evaluated using a
  regular mesh corresponding to $89^{3}$ points, $106^{3}$ points, and 
  $165^{3}$ points in the full BZ.  The $\bm{q}$-vector
  was varied along the [001] direction and the cone angle is
  \mm{\theta=20^{\circ}}.  The energies were calculated
  employing self-consistent potentials. }
\label{fig-ene-nktab}
\end{figure}

To obtain correct and unambiguous results on $D$ from spin spirals
energies, as outlined by Eqs.~(\ref{eq-D}) and (\ref{eq-ene}), it is
necessary to assess critically the fitting procedure which provides
the $D$ coefficient.  To find the power expansion of the energy
Eq.~(\ref{eq-ene}), the function $\epsilon(\bm{q})$ is fitted by a
polynomial via a least-squares method, within a certain interval of
\qq.  Therefore one has to check how the expansion coefficient $D$ is
affected by the degree of the polynomial which is fitted
to~$\epsilon(\bm{q})$ and also by the size of the interval within
which this fit is determined.

\begin{table}
\caption{Spin wave stiffness constant $D$ (in meV~\AA$^{2}$) of Fe
  evaluated by fitting the energy dispersion Eq.~(\ref{eq-ene}) by
  polynomials Eq.~(\ref{eq-polyn}) of degrees $2n$=2, 4, 6, 8, and 12,
  within intervals of different sizes. }
\label{tab-efit-Fe}
\begin{ruledtabular}
\begin{tabular}{cdddd}
  \multicolumn{1}{l}{polyn.} &
  \multicolumn{4}{c}{interval for fitting (in units of $2\pi/a_{0}$)} \\
  \multicolumn{1}{l}{degree} &
  \multicolumn{1}{c}{[0,0.05]} &
  \multicolumn{1}{c}{[0,0.10]} &
  \multicolumn{1}{c}{[0,0.15]} &
  \multicolumn{1}{c}{[0,0.20]}  \\
\hline
 4  &  301.0  &  302.6  &  302.5  &  316.8 \\
 6  &  303.6  &  302.6  &  298.6  &  298.1 \\
 8  &  298.6  &  301.6  &  305.0  &  294.3 \\
10  &  276.2  &  301.6  &  304.8  &  302.8 \\
12  &  251.6  &  302.1  &  300.6  &  302.8 \\
\end{tabular}
\end{ruledtabular}
\end{table}

Concerning the choice of the interval for the fit, there is the
natural requirement that the interval is not very large, because a fit
within a small interval emphasizes the behavior of \epsq\ at the
origin and that is what we aim at. However, for small $q$, there are
 technical 
problems with the numerical accuracy of the difference
$E(\bm{q},\theta)-E(0,\theta)$.
 Namely, if the number of $\bm{k}$-points increases, the energy
  dispersion curves $E(\bm{q},\theta)$ approach each other not
  uniformly but in a quasi-oscillatory way. 
This is illustrated in
Fig.~\ref{fig-ene-nktab} where the energy dispersion
\mm{[E(\bm{q})-E_{0}]/\sin^{2}\theta}\ for spin spiral waves in Fe
obtained using different $\bm{k}$-meshes is displayed very close to
the origin $|\bm{q}|$=0.  One can see that fine details of the energy
dispersion still vary even for quite dense meshes.  If the
$\bm{k}$-mesh density is not high enough, the behavior of the
\epsq\ function may significantly deviate from the expected form.
E.g., for the mesh with $106^{3}$ points, there is actually a local
{\em maximum} at $q$=0; it is shallow and can be seen only if the step
in $q$ is sufficiently small but it clearly hinders finding the
correct power expansion coefficients.  If one wants to by-pass the
numerical problems with determining $E(\bm{q},\theta)-E(0,\theta)$ for
small $q$ by performing the fit within a large interval, one has to
include more terms in the fitting polynomial, because as one moves
away from $q$=0, higher order terms get more important.  A proper
balance between the size of the interval in which the fit to \epsq\ is
performed and the order of the fitting polynomial thus has to be
achieved.

\begin{table}
\caption{As Tab.~\ref{tab-efit-Fe} but for Ni. }
\label{tab-efit-Ni}
\begin{ruledtabular}
\begin{tabular}{cdddd}
  \multicolumn{1}{l}{polyn.} &
  \multicolumn{4}{c}{interval for fitting (in units of $2\pi/a_{0}$)} \\
  \multicolumn{1}{l}{degree} &
  \multicolumn{1}{c}{[0,0.05]} &
  \multicolumn{1}{c}{[0,0.10]} &
  \multicolumn{1}{c}{[0,0.15]} &
  \multicolumn{1}{c}{[0,0.20]}  \\
\hline
 4  &  746.3  &  751.9  &  761.7  &  768.8 \\
 6  &  774.1  &  747.5  &  749.1  &  760.2 \\
 8  &  648.9  &  747.8  &  749.0  &  750.3 \\
10  &  463.9  &  747.2  &  747.4  &  747.6 \\
12  &  353.6  &  747.1  &  748.8  &  745.1 \\
\end{tabular}
\end{ruledtabular}
\end{table}

Tables~\ref{tab-efit-Fe}--\ref{tab-efit-Py} summarize the stiffness
constant $D$ for Fe, Ni, and Py evaluated by fitting the energy
dispersion Eq.~(\ref{eq-ene}) to polynomials of different degrees,
within intervals of different sizes.  We consider polynomials of even
powers only (because of the symmetry), they can be symbolically
written as
\begin{equation}
  f_{2n}(\bm{q}) \: = \: \sum_{i=1}^{n} a_{2i} q^{2i}  \; .
  \label{eq-polyn}
\end{equation}
The spin wave stiffness constant $D$ corresponds to the
quadratic term, 
\begin{equation}
  D \: = \: a_{2}
\; .
\end{equation}
The highest degree of a polynomial employed within this study is
twelve.  The spiral cone angle was set to \mm{\theta=20^{\circ}},
energies were obtained by means of self-consistent calculations, and
the $\bm{k}$-space integration was carried out on a mesh of $165^{3}$
points in the full BZ for Fe, $149^{3}$ points for Ni, and $133^{3}$
points for Py.

Inspecting Tabs.~\ref{tab-efit-Fe}--\ref{tab-efit-Py} gives an idea
about the stability of the procedure.  Fitting $\epsilon(\bm{q})$
within the smallest interval [0,0.05] (in units of \mm{2\pi/a_{0}}) is
clearly unstable, due to the problems with the $\bm{k}$-mesh
convergence (see also Fig.~\ref{fig-ene-nktab}).  When fitting within
larger intervals, one should employ polynomials of at least sixth
degree.  For the largest interval [0,0.20], the results sometimes
depend on the choice of the polynomial even up to the twelfth degree
(see Tab.~\ref{tab-efit-Ni}), suggesting that more complicated trends
which cannot be described by a simple polynomial may be present (see
the upper panels of Figs.~\ref{fig-eneq-fe}--\ref{fig-eneq-ni} for an
overall picture).  As a whole, however, for each of the systems one
can find a ``region of stability'' at the bottom right corner of the
respective table, where the values do not significantly depend on the
size of the fitting interval or on the degree of the fitting
polynomial
 (within the accuracy of $\pm$0.2~meV~\AA$^{2}$ determined by the
  $\bm{k}$-mesh convergence).  The numbers in this region do not
  depend on the fine details of how the coefficient at the quadratic
  term has been determined, therefore, they 
can be considered as the correct
artefact-free values of the stiffness constant $D$.  The spread of the
values within this region can be used to estimate the error of $D$ if
it is determined by fitting the spin wave energy dispersion.

\begin{table}
\caption{As Tab.~\ref{tab-efit-Fe} but for Py. }
\label{tab-efit-Py}
\begin{ruledtabular}
\begin{tabular}{cdddd}
  \multicolumn{1}{l}{polyn.} &
  \multicolumn{4}{c}{interval for fitting (in units of $2\pi/a_{0}$)} \\
  \multicolumn{1}{l}{degree} &
  \multicolumn{1}{c}{[0,0.05]} &
  \multicolumn{1}{c}{[0,0.10]} &
  \multicolumn{1}{c}{[0,0.15]} &
  \multicolumn{1}{c}{[0,0.20]}  \\
\hline
  4  &  534.4  &  519.8  &  525.5  &  528.4 \\
  6  &  505.2  &  518.9  &  521.0  &  524.5 \\
  8  &  342.9  &  518.0  &  521.6  &  521.4 \\
 10  &  205.7  &  499.0  &  521.5  &  520.6 \\
 12  &  170.6  &  535.4  &  521.2  &  520.9 \\
\end{tabular}
\end{ruledtabular}
\end{table}



\subsection{Weighted sum of $J_{ij}$ constants}

\label{sec-jij}

\begin{figure}
\includegraphics[viewport=0.2cm 0.5cm 8.7cm 11.0cm]{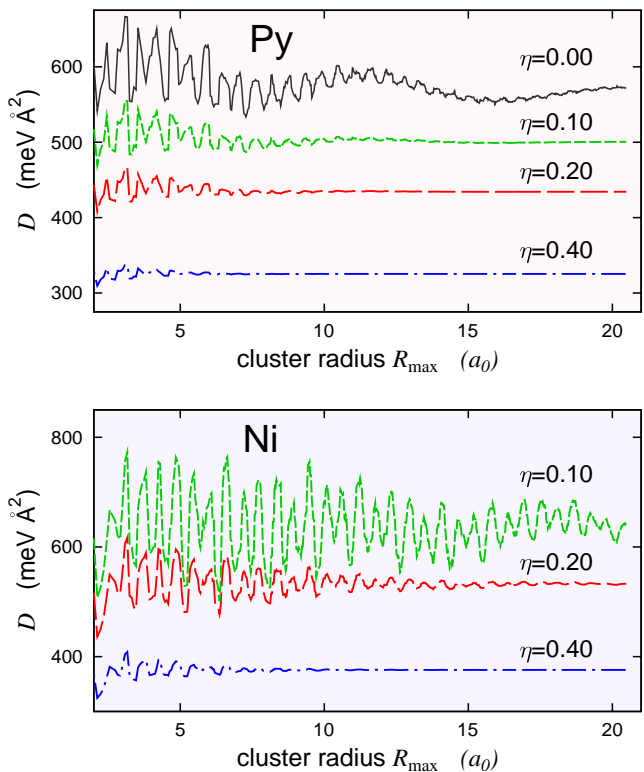}%
\caption{(Color online) Dependence of the spin wave stiffness constant
  $D$ on the maximum distance $R_{\text{max}}$ up to which the terms
  in Eq.~(\ref{eq-dsum}) are included, for different damping
  parameters~$\eta$.  Lower panel shows data for for Ni obtained using
  the mesh of $112^{3}$ points in the full BZ, upper panel shows data
  for Py obtained using the mesh of $75^{3}$ points in the full BZ.
  The distance $R_{\text{max}}$ is in units of the lattice constant
  $a_{0}$.}
\label{fig-DR-eta}
\end{figure}

In this section we inspect problems that may be encountered when
evaluating the stiffness constant $D$ via a weighted sum of the
coupling constants $J_{ij}$, as in
Eqs.~(\ref{eq-dlim})--(\ref{eq-dsum}).  Basic understanding can be
gained by looking on the dependence of $D$ on the maximum distance
$R_{\text{max}}$ up to which the individual terms in
Eq.~(\ref{eq-dsum}) are evaluated.  This is presented in
Fig.~\ref{fig-DR-eta} for Ni and Py, for several values of the damping
parameter~$\eta$.  One can see that the quasi-oscillations of
$D(R_{\text{max}})$ extend to quite large distances and that the
limiting value \mm{\lim_{R_{\text{max}} \rightarrow \infty}
  D(R_{\text{max}})}\ depends on the damping parameter~$\eta$.

\begin{figure}
\includegraphics[viewport=0.2cm 0.5cm 8.7cm 6.0cm]{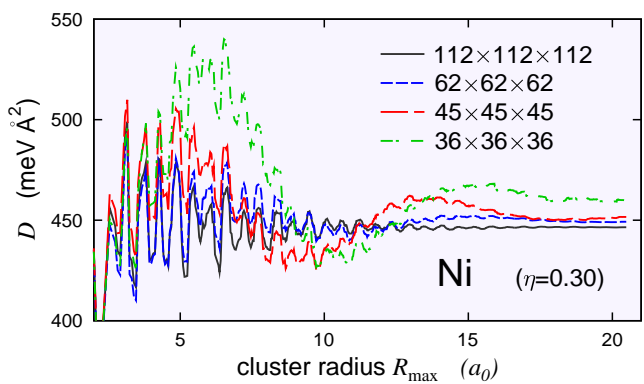}%
\caption{(Color online) Dependence of the stiffness constant~$D$ for
  Ni on the maximum distance $R_{\text{max}}$ up to which the terms in
  Eq.~(\ref{eq-dsum}) are included, for different numbers of
  $\bm{k}$-points in the full BZ (shown in the legend).  The data were
  obtained for damping parameter $\eta$=0.30. }
\label{fig-DR-nktab}
\end{figure}

The stiffness constant $D$ is finally determined via \mm{D=\lim_{\eta
    \rightarrow 0} D(\eta)}.  The limit has to be found by
extrapolating $D(\eta)$ down to $\eta$=0. Therefore, one should
evaluate $D(\eta)$ for as small~$\eta$ as possible.
Fig.~\ref{fig-DR-eta} demonstrates that to evaluate $D(\eta)$ for
small~$\eta$, one has to extend the sum in Eq.~(\ref{eq-dsum}) up to
large $R_{\text{max}}$.  Evaluating the exchange coupling constants
$J_{ij}$ for large interatomic distances requires a high density of the
mesh used for integration in $\bm{k}$-space \cite{PKT+01}.  To
illustrate this, we present in Fig.~\ref{fig-DR-nktab} the dependence
of the constant $D$ of Ni on the cut-off distance $R_{\text{max}}$,
for several $\bm{k}$-space grids.  It can be seen immediately that
going to larger $R_{\text{max}}$ requires a denser $\bm{k}$-mesh, 
 increasing dramatically the demand on conputational resources.

\begin{figure}
\includegraphics[viewport=0.2cm 0.5cm 8.7cm 11.0cm]{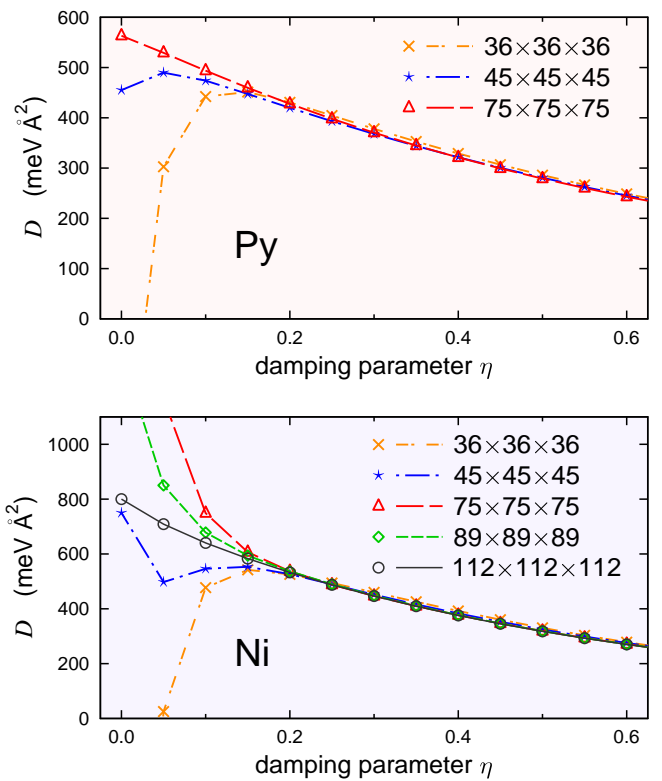}%
\caption{(Color online) Dependence of the spin wave stiffness constant
  $D$ for Ni (lower panel) and for Py (upper panel) on the damping
  parameter~$\eta$.  Data are shown for different $\bm{k}$-space
  grids. The maximum distance $R_{\text{max}}$ up to which the
  individual terms in Eq.~(\ref{eq-dsum}) were evaluated is
  20.5~$a_{0}$.}
\label{fig-etafit-nktab}
\end{figure}

The conclusion is thus the following: if we want to evaluate the
stiffness constant $D$ via Eq.~(\ref{eq-simple}), we have to
extrapolate $D(\eta)$ to $\eta$=0, which requires obtaining $D(\eta)$
for small~$\eta$'s, which requires extending the sum in
Eq.~(\ref{eq-dsum}) to large $R_{\text{max}}$ and that requires a high
density for the $\bm{k}$-mesh.  Graphically this is depicted in
Fig.~\ref{fig-etafit-nktab}, where we show how the stiffness constant
$D(\eta)$ of Ni and Py depends on the damping parameter $\eta$, for
several choices of the $\bm{k}$-mesh.  The summation in
Eq.~(\ref{eq-dsum}) covers interatomic distances up to
$R_{\text{max}}=20.5a_{0}$.  It can be clearly seen that if the~$\eta$
parameter is relatively large, the values of $D(\eta)$ do not depend
on the $\bm{k}$-mesh; the situation is numerically stable.  However,
for small~$\eta$, the values of $D(\eta)$ depend strongly on the
$\bm{k}$-mesh density.  Data for $\eta<0.2$ cannot be regarded as
numerically stable.

\begin{table}
\caption{Spin wave stiffness constant $D$ of Fe evaluated by summing
  the \mm{J_{0j}R_{0j}^{2}} terms,
  Eqs.~(\ref{eq-dlim})--(\ref{eq-dsum}).  The extrapolation of
  $D(\eta)$ to $\eta$=0 has been done by fitting $D(\eta)$ by a
  polynomial, within specific intervals of~$\eta$ values.  The
  interval within which the fit is done is specified in the first
  column, further columns contain values of $D$ obtained by employing
  a fitting polynomial of the second, third, and fifth degree in
  $\eta$. }
\label{tab-eta-fe}
\begin{ruledtabular}
\begin{tabular}{cccc}
  &
  2nd degree & 3rd degree & 5th degree \\
  \multicolumn{1}{c}{$\eta$} &
  \multicolumn{1}{c}{\mbox{$D$}} &
  \multicolumn{1}{c}{\mbox{$D$}} &
  \multicolumn{1}{c}{\mbox{$D$}} \\
  \multicolumn{1}{c}{interval} &
  \multicolumn{1}{c}{\mbox{(meV~\AA$^{2}$)}} &
  \multicolumn{1}{c}{\mbox{(meV~\AA$^{2}$)}} &
  \multicolumn{1}{c}{\mbox{(meV~\AA$^{2}$)}} \\
\hline
  0.2--1.0  &  269.7  &  278.1  &  281.6   \\
  0.4--1.0  &  261.7  &  275.7  &  276.8   \\
  0.6--1.0  &  252.9  &  273.0  &  296.7   \\
\end{tabular}
\end{ruledtabular}
\end{table}

The fact that the values of $D(\eta)$ obtained for low~$\eta$ are not
reliable questions the accuracy with which the stiffness constant can
be determined.  The extrapolation of $D(\eta)$ to $\eta$=0 is,
obviously, a delicate procedure depending on several technical
parameters.  This is illustrated in
Tabs.~\ref{tab-eta-fe}--\ref{tab-eta-py} where we show the stiffness
constant $D$ for Fe, Ni, and Py obtained by extrapolating $D(\eta)$ to
$\eta=0$ using different methods.  In particular, the extrapolation
was done using a polynomial of the second, third, or fifth degree in
$\eta$, determined by least-squares fitting of $D(\eta)$ when $\eta$
lies in the interval [0.2,1], [0.4,1], or [0.6,1].  The summation
Eq.~(\ref{eq-dsum}) includes all sites up to the distance
$R_{\text{max}}=20.5a_{0}$, which means about 70000~atoms (563
coordination shells) for bcc Fe and about 136000~atoms (773
coordination shells) for fcc Ni and Py. The $\bm{k}$-space integrals
needed to evaluate the $J_{ij}$ constants were carried out on a mesh
of $89^{3}$ points in the full BZ for Fe, $112^{3}$ points for Ni, and
$75^{3}$ points for Py.

\begin{table}
\caption{As Tab.~\protect\ref{tab-eta-fe} but for Ni. }
\label{tab-eta-ni}
\begin{ruledtabular}
\begin{tabular}{cccc}
  &
  2nd degree & 3rd degree & 5th degree \\
  \multicolumn{1}{c}{$\eta$} &
  \multicolumn{1}{c}{\mbox{$D$}} &
  \multicolumn{1}{c}{\mbox{$D$}} &
  \multicolumn{1}{c}{\mbox{$D$}} \\
  \multicolumn{1}{c}{interval} &
  \multicolumn{1}{c}{\mbox{(meV~\AA$^{2}$)}} &
  \multicolumn{1}{c}{\mbox{(meV~\AA$^{2}$)}} &
  \multicolumn{1}{c}{\mbox{(meV~\AA$^{2}$)}} \\
\hline
  0.2--1.0  &  708.3  &  754.6  &  768.1   \\
  0.4--1.0  &  663.2  &  736.7  &  769.1   \\
  0.6--1.0  &  618.2  &  712.9  &  785.5   \\
\end{tabular}
\end{ruledtabular}
\end{table}

One can see from Tabs.~\ref{tab-eta-fe}--\ref{tab-eta-py} that the
extrapolated values \mm{D(\eta\rightarrow0)}\ significantly depend on
the choice of the fitting interval.  The most conclusive estimates of
$D$ are those obtained using a polynomial fitted to $D(\eta)$ within
an interval which includes the smallest usable values for~$\eta$,
i.e., $\eta \in$[0.2--1.0].  Decreasing the lower boundary of the
fitting interval even further is not desirable because the values of
$D(\eta)$ may be numerically unstable for $\eta<0.2$ (see
Fig.~\ref{fig-DR-eta}).  As concerns the degree of the polynomial used
for the extrapolation: if the extrapolation is done via the second
degree polynomial in $\eta$, the outcome significantly depends on the
choice of the interpolating interval.  This can hardly be considered
as robust or stable.  On the other hand, for the fifth degree
polynomial, this dependence is only mild.  For the third degree
polynomial, the situation is somewhere in between. We can thus
conclude that estimating the stiffness constant $D$ by fitting the
$D(\eta)$ dependence by a fifth degree polynomial in $\eta$ leads to
trustworthy results.

\begin{table}
\caption{As Tab.~\protect\ref{tab-eta-fe} but for Py.  }
\label{tab-eta-py}
\begin{ruledtabular}
\begin{tabular}{cccc}
  &
  2nd degree & 3rd degree & 5th degree \\
  \multicolumn{1}{c}{$\eta$} &
  \multicolumn{1}{c}{\mbox{$D$}} &
  \multicolumn{1}{c}{\mbox{$D$}} &
  \multicolumn{1}{c}{\mbox{$D$}} \\
  \multicolumn{1}{c}{interval} &
  \multicolumn{1}{c}{\mbox{(meV~\AA$^{2}$)}} &
  \multicolumn{1}{c}{\mbox{(meV~\AA$^{2}$)}} &
  \multicolumn{1}{c}{\mbox{(meV~\AA$^{2}$)}} \\
\hline
  0.2--1.0  &  539.4  &  560.6  &  563.1   \\
  0.4--1.0  &  518.5  &  555.1  &  565.4   \\
  0.6--1.0  &  495.9  &  546.1  &  567.6   \\
\end{tabular}
\end{ruledtabular}
\end{table}

\begin{figure}
\includegraphics[viewport=0.2cm 0.5cm 8.7cm 6.5cm]{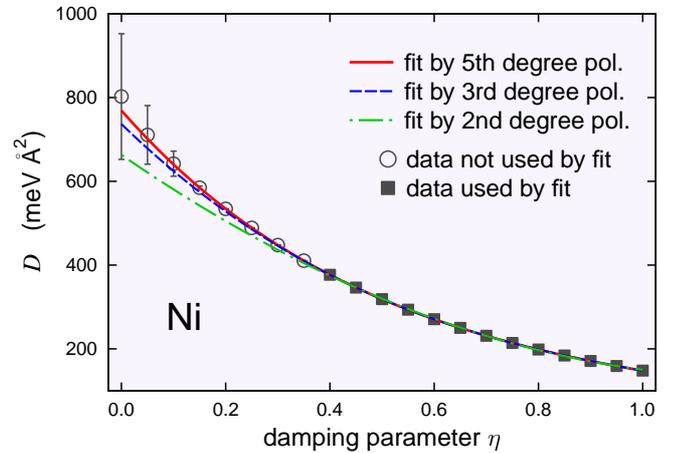}%
\caption{(Color online) Dependence of the spin wave stiffness constant
  $D$ for Ni on the damping parameter~$\eta$, together with fits of
  $D(\eta)$ by polynomials of the second, third, and fifth degree in
  $\eta$ (as indicated by the legend).  To obtain the fitting
  polynomials, only data for \mm{\eta \in [0.4,1.0]}\ were used.  The
  $\bm{k}$-space integrals were evaluated using the grid of $112^{3}$
  points in the full BZ, the maximum distance $R_{\text{max}}$ up to
  which the terms in Eq.~(\ref{eq-dsum}) were included is
  20.5~$a_{0}$.}
\label{fig-etafit-polyn}
\end{figure}

A complementary picture can be obtained by inspecting
Fig.~\ref{fig-etafit-polyn}, where we show the calculated values of
$D(\eta)$ for Ni, together with three different polynomial fits of the
$D(\eta)$ dependence. Note that if $\eta\leq0.10$, the values of
$D(\eta)$ contain significant numerical errors (see the bottom panel
of Fig.~\ref{fig-DR-eta}) --- we indicate this by errorbars.  The
polynomials of the second, third, and fifth degree in $\eta$ were
obtained by a least-squares fit for $\eta\in$[0.4,1.0].  One can see
that the quadratic fit (green dash-dotted line) fails to reproduce
$D(\eta)$ outside the fitting range --- it significantly deviates from
the values marked by open circles in Fig.~\ref{fig-etafit-polyn}.  The
situation is better for the cubic fit.  The best outcome is apparently
achieved if the fit is done by a fifth-degree polynomial.


\subsection{Direct evaluation of $D$ in reciprocal space}  

\label{sec-recip}

The third way to calculate the spin wave stiffness constant $D$ is via
integrating the relevant expression in reciprocal space, according
to Eq.~(\ref{eq-sergey}).  As this is a direct evaluation, no analysis
of the fitting or extrapolation procedure is needed.  The values we
obtained from Eq.~(\ref{eq-sergey}) are given in the fourth column of
Tab.~\ref{tab-compar}.



\section{Discussion}   \label{sec-discuss}

Our goal was to critically review different approaches to calculate
the spin wave stiffness constant $D$ and to compare the results
calculated for selected systems on the same footing.  The approaches
we investigated include (i) examining the long-wave-length behavior of
the spin wave energy dispersion [see Eqs.~(\ref{eq-D}) and
  (\ref{eq-ene}) and Sec.~\ref{sec-spir}], (ii) evaluating a weighted
sum of exchange coupling constants
[Eqs.~(\ref{eq-dlim})--(\ref{eq-dsum}) and Sec.~\ref{sec-jij}], and
(iii) direct evaluation of Eq.~(\ref{eq-sergey}) in reciprocal space.
The results for Fe, Ni, and Py obtained by these methods are
summarized in Tab.~\ref{tab-compar}.  The errors were estimated by
considering the $\bm{k}$-mesh convergence (for all three cases) and,
additionally, considering the ambiguity of finding the right fits for
$\epsilon(\bm{q})$ in case of analysis of the spin wave energy
dispersion (see Tabs.~\ref{tab-efit-Fe}--\ref{tab-efit-Py}) and of
extrapolating $D(\eta)$ to $D(\eta\rightarrow 0)$ in case of the
\mm{\sum J_{0j} \, R_{0j}^{2}}\ summation in real space (see
Tabs.~\ref{tab-eta-fe}--\ref{tab-eta-py}).  A minor contribution to
the errors comes also from the $\theta\rightarrow0$ limit when
determining $\epsilon(\bm{q})$ and from having a finite
$R_{\text{max}}$ when determining $D(\eta)$.
 These last two contributions are quite small in comparison with
  the errors due to the ambiguity of the fitting and/or extrapolating
  procedure.

\begin{table}
\caption{Overall estimates of the stiffness constant $D$ for Fe, Ni,
  and Py based on fitting the spin wave energy dispersion as in
  Eq.~(\ref{eq-D}) (second column), by performing a weighted sum of
  $J_{ij}$ constants as in Eq.~(\ref{eq-simple}) (third column), and
  by a direct integration of scattering matrices and operators in 
  reciprocal space as in Eq.~(\ref{eq-sergey}) (fourth column). }
\label{tab-compar}
\begin{ruledtabular}
\begin{tabular}{cccc}
  & \multicolumn{1}{c}{from $\epsilon(\bm{q})$}
  & \multicolumn{1}{c}{from \mm{\sum J_{0j} \, R_{0j}^{2}}} 
  & \multicolumn{1}{c}{from \mm{
      \int_{\text{BZ}} [T (\partial \tau/\partial k)]^{2}
    } } \\
\hline
Fe  &  302$\pm$2  & 279$\pm$2  &  262$\pm$3  \\
Ni  &  747$\pm$4  & 768$\pm$6  &  781$\pm$7  \\
Py  &  521$\pm$1  & 563$\pm$2  &  512$\pm$5  \\
\end{tabular}
\end{ruledtabular}
\end{table}

Small but distinct differences are evident in Tab.~\ref{tab-compar},
even though all the approaches use very similar physical assumptions.
In particular, in all cases it is assumed that the magnetization
direction can be described by vectors $\bm{\hat{e}}_{i}$ pinned to
atomic sites $i$  \cite{LKG84,LKAG87}. 
All calculations have been performed within the
same KKR Green function formalism, using the {\sc sprkkr} code,
 ensuring that the quantities used in different approaches
  (coupling constants, spin wave energies) are consistent.

To point out the differences in the approaches used here, we start by
noting that the scalar relativistic spin spiral calculations are
performed selfconsistently for each wave vector \qq.  As a
consequence, the exchange splitting of the energy bands as well as the
local magnetic moments are \qq-dependent. Moreover, the spin-spiral
energy has been evaluated based on the total energy of the system.
The other two approaches, i.e., the methods based on the real-space
summation \mm{\sum J_{0j} \, R_{0j}^{2}}\ according to
Eqs.~(\ref{eq-dlim})--(\ref{eq-dsum}) and on the reciprocal space
integral \mm{\int_{\text{BZ}} [T (\partial \tau/\partial k)]^{2}}
according to Eq.~(\ref{eq-sergey}), rely on the magnetic force theorem
and assume that the magnitude of the magnetic moments does not change
if they are tilted (rigid spin approximation).  These two approaches
are formally equivalent, as it was shown, e.g., by Liechtenstein
\ea\ \cite{LKG84}. However, small differences in the results occur
because the approaches lay different requirements concerning the
accuracy of numerical calculations. The requirements laid by the
real-space approach are discussed above in details. The approach based
on the Brillouin-zone integration is very sensitive to the features of
the electronic structure because of the $\bm{k}$-derivatives of the
$\tau$-matrix in Eq.~(\ref{eq-sergey}), in contrast to the real-space
approach. As a result, a very dense $\bm{k}$-mesh is needed for the
BZ-integration to achieve convergence with respect to the number of
$\bm{k}$-points.

The effect of using the magnetic force theorem can be seen from the
data in Tabs.~\ref{tab-spir-D-fe}--\ref{tab-spir-D-py}: it may result
in errors of few percents.  The assumption that the magnetic moments
do not change their magnitude if there are tilted is plausible for the
systems we are dealing with (see, e.g., Ref.~\cite{RJ+97} or lower
panels of Figs.~\ref{fig-eneq-fe}--\ref{fig-eneq-ni}), nevertheless,
small differences still may occur because of this.

\begin{table}
\caption{Experimental results for spin wave stiffness $D$ (in
  meV~\AA$^{2}$) for Fe, Ni, and  Fe$_{1-x}$Ni$_{x}$ alloy. Each study
  is identified by a 
  reference and the publication year.
   The concentration of Ni in 
  Fe$_{1-x}$Ni$_{x}$ is given in the last column.   } 
\label{tab-d-exp}
\begin{ruledtabular}
\begin{tabular}{rcccc}
  \multicolumn{1}{c}{work}  &  Fe & Ni &  Fe$_{1-x}$Ni$_{x}$
  & \multicolumn{1}{c}{$x$}  \\
\hline
\cite{HKL+64}\hspace{1em}(1964)  &
325   & 400   &  400 & 80  \\
\cite{Phi+66}\hspace{1em}(1966)  &
350   &   &  & \\
\cite{Str+68}\hspace{1em}(1968)  &
314   & 470   &  &  \\
\cite{MLN+73}\hspace{1em}(1973)  &
    & 555   &   & \\
\cite{Rie+73}\hspace{1em}(1973) &
311   &   &  &   \\
\cite{HHC+75}\hspace{1em}(1975)  &
   & 525 & 335 & 68 \\
\cite{Ald+75}\hspace{1em}(1975)  &
   & 555   &  &  \\
 \cite{MYW+76}\hspace{1em}(1976) &
   & 390   &  & \\
\cite{Rie+77}\hspace{1em}(1977)  &
   & 398   &   &  \\
\cite{LM+81}\hspace{1em}(1981)   &
   &  593   &  &  \\
\cite{Pau+82}\hspace{1em}(1982)  &
270   & 413   &  &  \\
\cite{Nak+83}\hspace{1em}(1983) &
   & 530  & 390 & 80 \\
\cite{LCL+84}\hspace{1em}(1984)  &
307   &  &  & \\ 
\cite{MP+85}\hspace{1em}(1985)   &
    &   398    &  & \\
\cite{YAD+17}\hspace{1em}(2017) &
 & & 440 & 80 \\
\end{tabular}
\end{ruledtabular}
\end{table}

As a whole, the differences between the values of $D$ obtained for
identical systems by different methods seem to be larger than what
could be ascribed to ``numerical noise''.  A closer look at
Tabs.~\ref{tab-efit-Fe}--\ref{tab-efit-Py} in Sec.~\ref{sec-spir} and
Tabs.~\ref{tab-eta-fe}--\ref{tab-eta-py} in Sec.~\ref{sec-jij} reveals
that use of just a bit different fitting and extrapolation method can
give rise to differences of 5--10~\%.  Even though we put a lot of
effort to compensate for ambiguities, some issues probably remained.
The third approach does not require any fitting or extrapolation but
it requires a very dense $\bm{k}$-mesh to get truly converged results
for the spin-wave stiffness, making the calculations very demanding.
We assume that the results obtained by means of Eq.~(\ref{eq-sergey})
could still be improved upon by increasing the $\bm{k}$-mesh density
but only at very high (impractical) computational costs.

On the other hand, possible errors of the procedures to calculate $D$
concern mostly the absolute values, not the trends.  For example,
application of the real-space summation
Eqs.~(\ref{eq-dlim})--(\ref{eq-dsum}) to Py doped with V, Gd, and Pt
led to a good theoretical description of the way the dopants influence
the spin wave stiffness, in agreement with experiment \cite{SME+19}.
Likewise, the dependence of the spin wave stiffness of the
Fe$_{1-x}$Ni$_{x}$ alloy on its composition can be properly described
both by fitting the spin wave energy dispersion according to
Eqs.~(\ref{eq-D}) and (\ref{eq-ene}) and by direct evaluation of $D$
via a reciprocal-space integral according to Eq.~(\ref{eq-sergey})
\cite{MPE+19}.

Uncertainty in determining the theoretical values of the stiffness
constants $D$ for Fe, Ni, and Py is accompanied by uncertainty in
experiment.  We summarize in Tab.~\ref{tab-d-exp} a selection of
available experimental data; a more complete list can be found, e.g.,
in Tables~7 and~8 of Vaz \ea\ \cite{VBL+08}.  One can see that the
spread of results of different studies is quite large --- about 15~\%
for Fe and Py and about 20~\% for Ni.  A critical assessment of
experimental studies is beyond our scope.  Despite the relatively
large spread of the data, Tabs.~\ref{tab-compar}--\ref{tab-d-exp}
indicate that our theory agrees well with experiment for Fe, whereas
for Ni and Py the agreement is less good.  Tentatively this is linked
to problems with describing the exchange coupling of Ni in terms of
the coupling constants \cite{Bru03,KDB08}.  Some errors could be also
introduced because of the restrictions of our computational scheme,
notably the ASA; nevertheless, full-potential effects are usually
small in close-packed metals such as those we are dealing with.  It is
more likely that the assumption of rigid moments is not fully
justified for Ni and its alloys.

 Even though our study has been performed for Fe, Ni, and Py, it
  is focused on analyzing and discussing concepts that have to be
  dealt with when studying the spin wave stiffness for any material.
  Metallic systems such as those we investigate here represent the
  most difficult case as concerns evaluating the stiffness by means of
  a weighted sum of the coupling constants (Sec.~\ref{sec-jij}).  This
  is because for metals the coupling constants $J_{ij}$ decay with
  distance as \mm{1/R_{ij}^3}\ \cite{PKT+01}, i.e., relatively slowly.
  For semiconductors and insulators, the opening of a gap leads to an
  exponential decay of the exchange coupling, as \mm{1/R_{ij}^3
    \exp(-R_{ij}/\lambda)}\ \cite{PKT+01,KTD+04}, improving the
  convergence of the expression Eq.~(\ref{eq-simple}) considerably.
  This will enable to employ larger $R_{\text{max}}$ and lower $\eta$,
  possibly disposing of the damping term $\exp(-\eta R_{0j}/R_{01})$
  altogether.  Increasing the disorder (as, for example, in the case
  of high-entropy alloys) will introduce an exponential spatial
  damping of the exchange coupling as well \cite{KTD+04}.

The analyses performed in Sec.~\ref{sec-res} enable us to draw some
recommendations how to evaluate the stiffness constant $D$.  In
general, evaluating $D$ by means of fitting $\epsilon(\bm{q})$ by a
polynomial is less demanding and more reliable than evaluating $D$ by
means of extrapolating $D(\eta)$ obtained by means of
\mm{J_{0j}R_{0j}^{2}}\ summation.  The convergence with the
$\bm{k}$-mesh density is better in the former case and, moreover,
extrapolation required in the latter case is always an ambiguous
procedure.  However, fitting \epsq\ by a polynomial to determine the
coefficient at the quadratic term is not without risks either.  Higher
powers should be included in the fitting polynomial
Eq.~(\ref{eq-polyn}); employing just a simple quadratic fit as done,
e.g., in \cite{SSL+17} may not always be sufficient.

In many cases, calculating the energy dispersion \epsq\ is difficult
or impractical (e.g., for multicomponent systems with substitutional
disorder).  Extrapolating $D(\eta)$ down to $\eta\rightarrow 0$ then
remains the only viable option.  In such cases, extra care has to be
taken and the robustness of the selected extrapolation procedure
should be checked.  For example, application of a quadratic
extrapolation within the $\eta\in$[0.6,1.0] interval (employed, e.g.,
for transition metals \cite{PKT+01} or for Heusler alloys
\cite{TCF+09}) to the systems investigated here would lead to a
systematic undershooting of $D$ by about 10~\%.  Of course, these
conditions have to be explored specifically for each system
considered.  Evaluating $D$ directly in reciprocal space
Eq.~(\ref{eq-sergey}) does not suffer from the pitfalls of fitting or
extrapolating but it is numerically demanding and sensitive to the
details of the electronic structure.


\section{Conclusions}   \label{sec-zaver}

Evaluating the spin wave stiffness constant $D$ by current schemes is
technically difficult and potentially numerically unstable.
Differences between values obtained by different methods of 5--10~\%
remain even if care is taken to make all the calculations consistent
with each other and well converged.  The agreement between theoretical
values and experimental data is good in case of Fe but significant
differences occur for Ni and permalloy Fe$_{0.19}$Ni$_{0.81}$.

Calculating the stiffness constant $D$ usually involves either fitting
the long-wave-length part of the spin wave energy dispersion \epsq\ by a
polynomial, or extrapolating the values $D(\eta)$ obtained via a
real-space summation of weighted exchange coupling constants to zero
damping, $\eta\rightarrow 0$.  Both procedures are tricky and quite
sensitive to technical details how the fitting or extrapolation is
done.


\begin{acknowledgments}
This work was supported by the GA~\v{C}R via the project 17-12925S and
by the Ministry of Education, Youth and Sport (Czech Republic) via the
project CEDAMNF CZ.02.1.01/0.0/0.0/15\_003/0000358. Additionally,
financial support by the DFG via Grant No.~EB154/36-1 is gratefully
acknowledged.
\end{acknowledgments}



\bibliography{liter-cube-and-py}

\end{document}